\documentclass[12pt]{article}
\begin{document}

\title{\bf Time inversion, Self-similar evolution and Issue of time}

\author{
Dhurjati Prasad Datta\\
Department of Mathematics \\
North Eastern Regional Institute of
Science and Technology\\
Itanagar-791109, Arunachal Pradesh, India 
\thanks{email:dpd@agni.nerist.ac.in}}
\date{}
\maketitle

\begin{abstract}
We investigate the question, "how does time flow?" and show that time may change by 
inversions as well. We discuss its implications to a simple class of linear systems. 
 Instead of introducing any unphysical behaviour, inversions can lead to a new multi-
time scale evolutionary path for the linear system exhibiting late time stochastic 
fluctuations. We explain how stochastic behaviour is injected into the linear 
system as a combined effect of an uncertainty in the definition of inversion 
and the irrationality of the golden mean number. We also give an ansatz for 
the nonlinear stochastic behaviour of (fractal) time which facilitates us to 
estimate the late and short time limits of a two-time  correlation function
relevant for the stochastic fluctuations in linear systems. These 
fluctuations are shown to enjoy generic $1/f$ spectrum. The implicit functional 
definition of the fractal time is shown to satisfy the differential equation 
$dx=dt$. We also discuss the relevance of intrinsic time in the present formalism, 
study of which is motivated by the issue of time in quantum gravity.
\end{abstract}


\newpage

\section{Introduction}
\par Time is one of the most enigmatic concepts in physics. Two of the most  
puzzling issues  in the present day theoretical physics: i)the issue of time in 
quantum gravity and ii) the ubiquitous presence 
of  $1/f$ spectra in diverse natural phenomena, though seemingly 
unrelated, are actually related to time.  As it is well known,  the 
timelessness of Wheeler-Dewitt equation, the quantal evolution equation  of a 
closed gravity-matter system, leads to longstanding conceptual issues in 
quantizing a gravitational system [1]. One of the key problems here is to give 
a realization of an intrinsic (internal) time,  which will allow  a  self-consistent 
description of the intrinsic relative changes of the interacting degrees of 
freedom, although the dynamics of the total system appears `frozen' from 
the point of view of the external time [2]. The issue of time is thus a mystery 
related to the quantal aspect of a closed system in the small time limit.  The origin of 
$1/f$ spectrum, on the other hand, relates to self-similar fluctuations in  a 
nonlinear system over very long time scales [3]. Although, much work have been 
done in this area over the last two decades [4],  a re-examination of 
the generic $1/f$ spectrum  in dynamical systems throwing new light into the problem 
would always be welcome. Even as the two issues are treated separately in 
the literature, we show, in the following, that these problems might have a common origin
in an extended dynamical framework incorporating time inversion as yet another physical 
mode of time increment. The study will not only provide a general explanation of the 
generic $1/f$ spectrum, but also lead to fundamentally new insights into the structure of 
time. To explain our basic premise (framework) we treat here only a simple 
class of evolutionary equation in classical mechanics. Applications to a genuine quantum 
gravity model would be considered separately. As will become clear, some new insights 
into the issue of intrinsic time would be gained even at this level of our presentation.

\par To motivate the extension of the ordinary dynamical framework, we 
begin by posing the following two closely related questions: i) How 
does time flow? ii) Can a linear system support self-similar structures 
(fluctuations)? Obviously, answer to the second question is `no', in the ordinary 
dynamics. Time (for that matter, any physical quantity (real variable)) is 
supposed to change (appart from trivial rescalings) by pure translation only. 
In the present paper, we however, advocate that time may change even by 
inversions, leading, in particular, to self-similar evolutions (flucutations) 
even for a linear system. The definition of time inversion has an {\em inherent} 
uncertainty, injecting stochastic behaviour in the late time evolution of the system. 
In the short time limit, on the other hand, this stochasticity could be exploited 
to derive a timeless Wheeler-Dewitt-like equation for the purely fluctuating 
component of the system variable. Although a self-consistent realization of  
intrinsic time is available {\em naturally} in the present simple model, its extension 
to a quantum gravity model may have to surpass  nontrivial problems. However, we 
believe that the present analysis would surely be of much utility in future 
studies of quantization problems of the gravitational fields.

\par It is worth pointing out here the relevance of the present work with 
recent studies. The possible presence of self-similar fluctuations/
dynamics in a linear system has been reported [5-7] recently. Bramwell et al [5]
showed that a linear spin-wave theory of a critical ferromagnet could
describe the fluctuation statistics of a closed  turbulance experiment.
Subsequently, a non-gaussian probability density function is derived [6],
which turns out to represent an `effective universality class'  for a large
number of strongly correlated systems. A common feature of these dissimilar
systems
is the presence of self- similar fluctuations over many length scales leading 
to fractal properties. It is also noted that the explicit nonlinearity may 
not have an essential role in generating these non-gaussian fluctuations.
In ref.[7], on the otherhand, we pointed out the presene of self- similar
fluctuations over all time scales in a time dependent quantal evolution. 
As a consequence, the late time behaviour of the evolving state would exhibit 
non- gaussian fractal features described by a fractal (uncertainty) exponent 
$\nu=(\sqrt 5-1)/2$, the golden mean. The analysis makes use of an intrinsic 
sense of time [2], which is introduced in the evolution exploiting the
emergent nonadiabatic geometric phase of the evolving state. As remarked above, 
the intrinsic time is meaningful in a closed system when the sense of time is 
defined internally from the relative changes of the interacting degrees of freedom
(observables) of the system [2], in contrast to the 
externally defined Newtonian time in ordinary dynamics (of open systems). 
It turns out that the intrinsic time, which tracks the evolution of the purely 
fluctuating state when the mean dynamical evolution of the total state is 
removed, relates inversely with the external Newtonian time $t$, endowing an 
SL(2,R) form-invariance to the corresponding Schr\"odinger equation.
Consequently, a fluctuation that is small in comparison to the mean 
evolution of the state in the scale $t\sim 1$, can be mapped to a large 
fluctuation (comparable to the mean ) in the scale of $t_1=\nu t\sim 1$  and 
is described analogously by an approximately self- similar Schr\"odinger 
equation, when $t_1$ acts as the time parameter. Contributions from all 
scales then lead to a nonvanishing exponent in the limit $t\rightarrow 
\infty$.

\par Clearly, the present study is in continuation of Refs.[2,7]. 
To understand the origin of a possible time inversion and
the associated self-similar evolutions in a 
time dependent {\em linear} evolution more clealy, we  study here a linear 
differential dynamical system in one space dimension, which may be considered 
as the simplest classical analogue of the Schr\"odinger equation. We show 
that even in the absence of a geometric phase, time inversion can be 
injected into a linear (classical) system by using the golden mean partition of 
unity: $\nu +\nu^2=1$. The meaning of time inversion and its 
relevance to linear systems is explained in Sec.2, by pointing out an intrinsic 
uncertainty in the definition of inversion. 
In Sec.3, we demonstrate how time inversions, in 
the context of the ordinary time, can lead to a  new  class of self-similar solutions 
to the linear equation. The origin of intrinsic time in the classical context is also 
pointed out. In Sec.4, we interprete the new solution as representing 
the late time stochastic behaviour (fluctuations) (over multiple time scales) of the 
standard solution, when randomness is injected into the system via time inversions 
at randomly distributed `transition moments', transfering the evolution from one scale 
to other. This random multiscale evolution of a linear system thus reveals a fractal- 
like aspect in time itself. We also discuss the relation of the inversion induced 
uncertainty to the irrationality of the golden mean, giving  rise to a fundamental 
limitation on measurability. In Sec.5, we present an ansatz for this stochastic 
fractal time, and discuss how this ansatz could successfully retrieve the late time 
stochastic behaviour of the linear system in a concrete way. We also point out 
here the close resemblance of the present stochastic equations with quantal 
evolution. The derivation of $1/f$ spectrum and applications to certain natural 
processes are discused in Sec.6. In Sec.7, we show that our nonlinear ansatz for the 
fractal time $T$ represents a new class of fractal solutions for the equation 
${dT\over dt}=1$. The role of the measurement uncertainty in 
formulating the fractal time concept is further highlighted here. 
We sum up our presentation in Sec.8. It will be clear that the discussion of Secs.2-4 
forms the background of the stochastic formalise of Secs.5-7.

\section{Time Inversion}
\par Let us consider the simplest linear dynamical system given by

$$
{{dx\over dt}=(1+v (t))x}\eqno(1)$$

\noindent where $t$ denotes the dimensionless (Newtonian) time.( We scale
$t$ suitably to adjust the dominant scale of evolution to $t\sim 1$).                                                            
The above equation is form-invariant under time translations. In ordinary
treatment, leading to the standard solution $x_s\sim \exp(t+
\int vdt)$, the variable $t$ (spacelike or timelike)
behaves only as a labeling parameter. The above equation, being linear
and deterministic, can not generate self-similar fluctuations
in the ordinary framework.
We, however, solve eq(1) as a true evolutionary
process, by following the evolution (unfolding) of the system $x$ as time
changes (flows) successively in steps of a given unit. We also look for
clues to extend the time translation form-invariance of eq(1) to the SL(2,R) 
form-invariance. We note that SL(2,R) represents the minimal extension
of the translation group incorporating inversions. 

\par Let us assume that the system given by eq(1) evolves
out from the initial time $t=0$ with $x(0)=1$. 
The (implicitly time dependent) function $v(t)$ may be considered to represent
the time dependent interactions of the environment on the system $x$. (This 
representation, though useful for our later discussion on intrinsic time, 
is not essential.)
Following the Born-Oppenheimer ansatz (and using the terminology of stochastic
processes, to be justified a posteriori), the evolution of the 
system can now
be decomposed into two parts $x=\tilde x x_1$, where $\tilde x=e^t$ is the
explicitly time dependent `mean' evolution and $x_1$ is the implicitly time
dependent purely `fluctuating' component satisfying the reduced equation

$${{dx_1\over dt}= v(t)x_1}\eqno(2)$$

\noindent We note that ordinarily a change in time, in the vicinity of a given 
instant $t_0$ is indicated by a pure translation $t=t_0+\bar t\equiv t_0+(t-t_0)$. 
In fact, the last equality is an identity (valid for all $t$). In the 
neighbourhood of $t=1(t>1)$ (the scale of explicit time dependence) (say) an 
increase of time is thus indicated by $t=1+\bar t$, $\bar t
=t-1\approx (>)0$. The standard solution, $x_s$, is clearly the unique translation 
form-invariant solution of eq(1). By inversion, on the otherhand, one means 
$t=(1+\bar t)^{-1}\approx 1-\bar t$.  However, 
contrary to pure translations, the inversion leads to an equation (in $t$),
fixing $t=1$, for $\bar t>0$.  Of course, one can consider the inversion as
the definition for the variable $\bar t$ itself. In that case, the inversion
turns out to be a trivial representation of the translation $t=1-\bar t,
\bar t=1-t$ near $t=(<)1$. Consequently, the possibility of an inversion is
ordinarily ruled out. In the following we, however, show that a nontrivial
realization of inversion (at the level of an identity) is indeed possible
in the vicinity of $t=1$ with interesting physical implications
in the context of an evolutionary equation of the form eq(1).

\par To this end, let us note that one is indeed free to interprete inversion
as a two-time tranformation. Let $t_{\pm}$ denote times $t>1$ and
$t<1$ respectively. Then close to $t=1$, the inversion $t_{-}=1/(1+(t_{+}-1))$
leads to the {\em constraint} $1-t_{-}=t_{+}-1$. The parametric representation
of inversely related times is obviously given by $t_{-}=1-\bar t$ and
$t_{+}=1+\bar t$, $0<\bar t<<1$ (so that the constraint reduces to an {\em 
identity}, valid close to $t=1$). With this reinterpretation, time inversion
in the vicinity of $t=1$ acquires the status of a pure translation.
If translation is considered to be the most natural mode of time increment, 
then there is no compelling reason of ignoring inversion
as yet another {\em natural} mode of doing this. Consequently, it seems 
reasonable to assume that time changes from $t_{-}$ to $t_{+}$ not only by
ordinary translation over the period $t_{+}-t_{-}=2\bar t$, but also 
instantaneously by inversion. 

\par To clarify the point further, we note that the above definition gives 
a {\em new nontrivial} solution $t_{-}=1/t_{+}$ to the constaint
$t_{-}+t_{+}=2$, in the {\em vicinity} of $t=1$ over the linear solution 
$t_{-}=2-t_{+}$, ordinarily thought to be the only possible solution. 
The reason for this belief obviously relates to the idea that 
a change in a variable $t$ (say) is accomplished only by pure translation.
Consequently, one may argue that the parametric form of our two-time (point) 
solution being clearly linear close to $t=1$, is a trivial reprsentation 
of the standard (linear) solution and must be devoid of any new dynamical
content.
Our intention, however, is to emphasize the contrary; although apparently 
linear, the two-time parametric solution, being inversely related, 
does indeed hide a structure of nonlinearity in the form of SL(2,R) group 
action, giving rise to an alternative means of inducing a change in the 
variable $t$. In view of this nonlinear possibility, the change (flow) of 
time could be visualized as an SL(2,R) group action, when the nontrivial
SL(2,R) action is realized only near $t=1$. Accordingly, time flows from 
$t=0$ to $t=t_{-}$ by translation, and then may switch over to $t_{+}$ by 
inversion $t_{+}=1/t_{-}$, for another period of linear flow etc. In Sec.3
we explain in more detail how this scenario gets a natural application in 
the context of eq(1) leading to new dynamical features. The restriction on the 
applicability of the nontrivial generator of SL(2,R) (viz., the inversion),
to be close to $t=1$ is vital. Once this is removed, the inversion solves the 
constraint only for $t_{-}=t_{+}=1$. Consequently, the new dynamical features 
would arise mostly in connection with the short time (viz., close to the 
moment $t=1$) and/or late time (by inversion) behaviour of the evolving system.
We note incidentally that the above definition
of inversion has an {\em inbuilt} uncertainty. The exact moment when an
inversion is materialised, is rather irrelavent; it can be at {\em any} instant
close to $t=1$. (To put it in another way, the nature of the `inversion 
constraint' allows the parameter $\bar t$ to be a random variable (c.f., Sec.5)). 
As will be pointed out below, this uncertainty is actually
related to the {\em irrationality} of the golden mean $\nu$ and is also
responsible for a loss of late time predictability
even for the simple system eq(1). 

\par Before closing this section, it is worthwhile to compare the present 
definition of time inversion with the usual time reversal (inversion) symmetry 
of an equation of the form eq(1). The usual time reversal symmetry means that 
the system $x(t)$ evolves not only forward in time from $t$ to $t+h,h>0$, 
but it can also evolve backward; i.e., the state $x(t)$ can be reconstructed from 
the state $x(t+h)$. The parameter $t$ in eq(1) is thus `non-directed', 
(cf. the remark below eq(1)) giving rise to the problem of time asymmetry  
( note that all the fundamental equations in Physics are time (reversal) 
symmetric). This is to be contrasted with a (time asymmetric )diffusion equation, 
having a well-defined temporal sense. We show below that the present definition of 
(time) inversion introduces a well-defined time-sense in the system's evolution 
following eq(1), analogous to a diffusive process. 
More precise derivation of this equivalence will be given elsewhere.

\section{New Solution}
\par To discuss the salient features of time inversion in the context 
of ordinary time $t$, it suffices to
restrict the class of equation (1) to the one given by $v(t)=v_0 t$.
The parameter $v_0$ may be a slowly varying function of $t$ so that
$|t{dv_0\over dt}|<< v_0$.
For simplicity, we may thus fix it to a (small) constant.
(The $t$ dependence of $v(t)$ thus gets explicit. However, the slow implicit
$t$ dependence of $v_0$ will be kept in view in the later discussion of
intrinsic time). One thus considers the `linear regime' of the system
evolution, nonlinear time dependence (interactions), if any,
would be felt only after an elapse of time $O(v_0^{-1})$.  Let us now note 
that a translation  near $t=1$, treated as a transformation between two 
independent variables $t$ and $\bar t$,
can lead to a scale changing transformation of the form 

$$
{t={{1+\nu t_1}\over {1-t_1}}}\eqno(3)$$

\noindent  where $t_1=\nu (t-1)\approx (>)0$. Indeed, for $t\approx(>) 1$ 
eq(3)  reduces, upto linear terms, $t\approx 1+\nu(\nu +1)(t-1)$, so that, 
$\nu(\nu+1)=1$. Thus, for $\nu=$the golden mean, eq(3) 
gives a nontrivial SL(2,R) representation of the pure translation.
We note that in the context of pure translations, one is unable to interprete 
(utilize) an equation of the form eq(3). To explore the
implications of  eq(3) in relation to an inversion, it is advantageous to rewrite 
$v_0$, without any loss of generality, as $v_0=\lambda^2\nu(>0)$. Moreover, we rescale 
$t$ by $\tilde t=\lambda t$ in the reduced equation, to rewrite it as 

$$
{{dx_1\over d\tilde t}=\nu \tilde t x_1}\eqno(4)$$

\noindent (We drop  tilde henceforth and reintroduce it later to discuss
the scaling properties of $x$. The parameter $\lambda$ would indicate
the strength of the nonlinear influence of time on the system.) To recall, 
eq(4) is derived by removing the 
mean $\tilde x$ of the total system after an elapse of time $t=t_{-}$ (say, for 
definiteness). By the inversion constraint $1-t_{-}=t_{+}-1$ the fluctuation 
$x_1$ is transported instantaneousely to $t_{+}$, since $t_{-}$ and $t_{+}$ 
are identified, so to speak, by inversion $t_{+}(t_{-})=t_{-}^{-1}=1+\bar t$. 
Noting $dt_{-}=-dt_{+}=-d\bar t$, and using eq(3) we get 
 
$$
{-{dx_1\over dt_1}=(1+\nu t_1)x_1}\eqno(5)$$

\noindent where $t_1=\nu\bar t\approx (>) 0$. Thus as the total system 
evolves upto $t_{-}\approx 1$ (say), the SL(2,R) representation eq(3) in 
conjuction with an explicit time inversion, 
restores the initial linear evolution to the fluctuation $x_1$ in time $t_1$. 
To emphasize, in the absence of time inversion, eq(5) would have been
impossible. Eq(3) is valid only in the vicinity of $t=1$, triggering
the transition of
eq(1) to eq(5) via eq(4), when eq(5), being a self-similar replica of eq(1) 
is valid at least upto $t_1\approx 1$. Consequently, as time flows from 
$t_{+}$ onward linearly by translation, the fluctuation $x_1$ now evolves 
under eq(5) till $t_1\approx 1$. The (-) sign is a nontrivial signature of 
time inversion. To understand its origin, we rederive eq(5) when the 
subtraction of the $\tilde x$ is accomplished at a time  $t_{+}(>1)$.
Following the above steps, eq(4) now gets transformed to 

$$
  {{dx_1\over dt_1}=(1-\nu t_1)x_1}\eqno(6)$$

\noindent when the inverted form of eq(3) is used to indicate the transfer 
of the fluctuating system from $t_{+}$ to $t_{-}$. One then needs to replace 
$t_1\rightarrow -t_1 (t_1>0)$ to put eq(6) in the form eq(5). The need for 
a sign change in $t_1$ is necessary to counter the backward time flow 
generated by the transition $t_{+}\rightarrow t_{-}$. The small scale 
{\em intrinsic} evolution of the fluctuation $x_1$ given by eq(5) is thus
realized in the opposite direction of the usual external time evolution
eq(4) [7]. In other words, the possibility of a backward flow in time is
avoided by flipping the direction of the  self-similar evolutions of the
system at successive iterates. To justify 
the term intrinsic, we note that the self-similar evolution in 
time $t_1$ is generated by splitting, so to speak, the time $t$ itself into 
the inverted time in the form of the factor $(1-t_1)^{-1}$ and the time 
($t_1$) dependent interaction  in eq(3). The self-similar eq(5) is thus 
a signature (and consequence) of the {\em self-interaction} due to the nonlinear 
structure in time, endowed by inversions.

\par To clarify it further, let us rederive eq(5) by yet another route, 
making the relationship between time inversion, intrinsic sense of time [2] 
and self-similar evolution more transparent. We start again from 
eq(4) but assume that the time of removal of $\tilde x$ is $t_{-}=1-
\bar t_1(\bar t_1>0)$, instead of $t_{-}=1-\bar t$, as in the previous cases.  
(To keep our notations clear, we denote a linear variable $t$ near $t=1$ by 
$t=1+ \bar t$, where $\bar t$ represents $t$ near $t=0$.) We note that the small 
scale time $t_1$ is defined intrinsically by the interaction, which, in turn,
introduces a squeezing of the ordinary time increment near $t=1$: $\bar t_1=
v(t)=\nu \bar t (\approx 0), 0<\nu<1$. The possibility of two choices of 
$t_{-}$ could be ascribed to the uncertainty in the exact moment injecting 
an inversion into the system near $t=1$. The proximity of the present moment 
to $t=1$ over the other is indicated by the scale factor $\nu$.
Noting that $t=\bar v(t_1)t [\bar v(t_1)]^{-1}$, $\bar v(t_1)=1+\nu \bar t_1$, 
we rewrite eq(4) as 

$$
{-{dx_1\over dt_{-}}=\bar v (\nu t_{-}) \bar v^{-1}x_1}$$
 
\noindent This equation now leads to eq(5) provided two moments $t_{-}$ near 
$t=1$ and $t_{+}$ near $t\sim\nu^{-1}$ are identified by inversion $t_{+}-
\nu^{-1}=1-t_{-}= \bar t_1$, i.e., when $\nu t_{+}=\bar v\approx 
1+\nu \bar t_1$ ($dt_{+}=-dt_{-}=d\bar t_1$). The factors $\bar v, \bar v^{-1}$ 
(being the small scale replica of the interaction term),
are introduced self-consistently from the intrinsically available 
informations (observables)[2] in the evolving fluctuation, eq(4), eliminating 
the external time variable $t$ of eq(1). This also removes any arbitrariness 
in the choice of the factors. Stated otherwise, the self-similar eq(5) may be 
considered to be an outcome of the self-measurement of $x_1$ by itself. 
In fact, the variable $\bar t_1$ indicates the time recorded, as it were,
by an `internal clock'
stationed in the total system $x$ itself, whose rate of variation is correlated 
with (and determined by) the variation of the interacation via
$v$, an implicit function of time $t$ (recall the slow time dependence
of $v_0$). (Thus, in relation to eq(5), the system $x$ has the role of the
`universe'[2,7].) Note that eq(4) is an extrinsic equation, since the changes in
$x_1$ there is measured by an external clock.  The inversion now carries the moment
$t_{-}$ to the smaller scale $t_1$ near $t_1=(>)1$ via $\nu (1-t_{-})=\nu t_{+}-1
\equiv t_1-1$, where $t_1=1+t_2,\,t_2=\nu^2\bar t$. Consequently,  
as the linear Newtonian time changes from $t=1$ onwards as $t=1+\bar t$, a 
concommitant flow of an intrinsic sense of time also gets developed in the 
system, which changes by inversion $t_{-}(=1-\bar t_1)=(1+\bar t_1)^{-1}$, 
near $t=1$, incorporating an explicit change of scale, leading to self- 
similarity, in the subsequent evolution over the period $t_1\approx 0$ to 
$t_1\approx 1$, and hence  relating smaller scales near $t=1$,
successively,  to longer scales as $t\rightarrow \infty$. This indicates 
how an otherwise uni-scale evolution of $x_1$ acquires multiscale self- similarity 
under inversions.
\par To summarize, the time inversion as defined in Sec.2 along with
the SL(2,R) representation, eq(3), can 
induce a self-similar evolution eq(5)
to the fluctuation $x_1$ when the mean evolution $\tilde x$ is removed near
$t\approx 1$. Time in the self-similar  eq(5) is recorded,
as it were, by a `clock' stationed intrinsically in the total system $x$,
so that the (intrinsic) time is measured in the unit of $\nu$, against the
external flow in the unit of $t\sim 1$. We remark that 
the removal of the mean $\tilde x$ can be accomplished at any
instant in a neighbourhood of $t=1$. This {\em uncertainty}, however, is not
apparent in the self-similar eq(5), as it is obtained via inversion at a well 
-defined instant, $t_{-}$ say, near $t=1$. In any case, the possibility of an 
instant like $t_{-}$ (and/or $t_{+}$) to act as a random variable is sufficient to 
inject a randomness in the late time evolution of the system. This point will 
be elaborated further in later sections.

\par Returning to the main discussion,  we note that eq(5) is valid upto
$t_1\approx 1$ ($t\approx 1+\nu^{-1}$),
when (the linear regime) of eq(1) is valid upto $t\approx \nu^{-1}$. Near 
$t_1=1$ the system $x_1$ again makes a transition to the 2nd order
self-similar fluctuation $x_2: x_1=e^{-t_1}x_2$ satisfying the equation

$$
{{dx_2\over dt_2}=(1+\nu t_2)x_2}\eqno(7)$$

\noindent and so on, to finer and finer scales as $t\rightarrow \infty$. 
The rate of this successive transitions is very {\em slow}. Indeed, as time 
$t$ flows linearly to $\infty$, the inversions at the successive transition 
moments $\tau_n=\Sigma_0^n {1\over \nu^n}$ generate, so to speak, an 
(overall) intrinsic flow of time: $T={1\over {1+t_1}}, t\approx 1$,
$T=[1,1;1+t_2]$ at $t\approx 1+\nu^{-1}$ so on, so that $T\rightarrow 
\nu$ through the sequence of 
the golden mean approximants as $t\rightarrow \infty$, the rate of whose 
convergence is the slowest possible. Note that after first iteration, $t_1$ 
in eq(5) flows from $t_1\approx 0$ to $t_1\approx 1$ linearly, thus making a
room for the 2nd iteration at $\tau_1$ and etc. The new intrinsic sense of
time denoted $T$ thus resembles, as it were, a cascaded flow down the ladders
of the golden mean continued fraction, transporting the evolution of the
system $x$ successively to scales $t_{n+1}$ at the transition moments $\tau_n$.
As pointed out already, the transition moments are inherently random, thus
inducing a {\em stochastic} nature in the intrinsic time $T$.
We recall that an iteration always generates a sense of time in the context
of a discrete dynamical system (map). The present scheme of iteration via
inversions, however, relates to two senses of temporal directions: the
intrinsic time sense $T$ which converges to $\nu$ and the ordinary linear
sense of $t\rightarrow \infty$, as $n\rightarrow\infty$. In any case, the
system  $x$, however, evolves {\em uniquely} along the intrinsic time  flow, 
thus traversing finer and finer scales $t_n$, as $n\rightarrow\infty$.
The scales $t_n$ can be treated as independent variables related to each
other successively by scaling equations $t_{n+1}=\nu (t_n-1)$  near 
$t_n=1$, $t_{n+1}=0$. In the limit $t\rightarrow \infty$, the  (intrinsic ) 
solution of eq(1) incorporating equal contributions from all scales thus has  
the form 
 $$
 {x_i=e^{(t-t_1+t_2-\cdots)}}\eqno(8)$$

\noindent where $t>t_1>t_2>\ldots $. We note that the randomness in $\tau_n$ 
makes the variables $t_n$
a set of stochastic variables, with randomness concentrating near $t_n=0$ and
$t_n=1$. The contributions from these
infinite number of (random) scales would  of course lead to a very complicated 
fractal -like structure for $x$. A somewhat simplified (but, nevertheless,
useful) view of this
fractal behaviour emerges when we look for  the late time (t) asymptotic 
form of eq(8) using the scaling relations $t_n= \nu^n t$ (c.f., our actual
derivation of eq(8) following system's evolution from $t=0$ to $t=\infty$
via inversions at well-defined transition moments and stretching each of the 
scales $t_n\rightarrow \infty$ via scaling relations). 
We get $ x_f \sim \eta^\nu$, where, $t_1=\ln\eta^{-1}, \eta\rightarrow 0$,
and $x_f =x_i/\tilde x$ stands for the renormalized form of $x_i$ when
the initial mean evolution (of the zeroth iterate) is subtracted out. 
We note incidentally that, in the absence of time inversion, the standard solution 
of eq(1), with a linear $v(t)$,  has the form $x_s=e^{t+{1\over 2}\nu t^2}$. 
To pave the discussions of the subsequent sections, 
we compare, in the following section, the intrinsic solution (8) with the standard 
solution $x_s$, and give an heusristic interpretation of  $x_f$ as a correlation 
function.

\section{Interpretation}
To interprete the new solution $x_i$, let us proceed in several steps. We
begin by noting that under inversion (i.e., the extended SL(2,R) symmetry)
the points $t=0$ and $t=\infty$  are identified. In the present context this
means that the system would enjoy identical (equivalent) dynamical
properties ( though their interpretations might vary, see below) for very
short and late time. As a check, one can easily verify that eq(1) with $v(t)=
\nu t, t=t_0(1+\eta), t_0$ large, gets mapped to eq(5), under inversion, when  
$\tilde t=t_0\eta=(1+\bar t), \eta\approx t_0^{-1}\approx 0$ and $\tilde x=
e^{(1+\nu t_0)t}$ (c.f., eq(4)).
We note next that the quadratic term in the exponential
of $x_s$ due to the linear time dependent term $v(t)\propto t$ in eq(1), is
replaced by the infinitely many scale dependent linear terms in the new
SL(2,R) invariant solution (8). The evolution  thus remains `truely' linear
over every time scale, in the sense that  even the  solution $\ln x_i$ is  
linear, though in multiple (random) scales (variables). The reason of this,
as pointed out  already, is that $v(t)$ continues to replicate using the 
`golden mean splitting ansatz' eq(3) down the scales, thus driving 
the nonlinear 
(quadratic) contribution out to infinitely distant time, by keeping it 
always insignificantly small.  Consequently, the solution (8) could as 
well be considered as the one valid in the short time limit near $t=1$. 
In fact, this follows  from the set of self-similar equations

$$
{{dx_n\over dt_n}=(-1)^n(1+\nu t_n)x_n}\eqno(9)$$

\noindent where $t_0\equiv t$, when the variables $t_n$'s are allowed to vary
in the finite interval (0,1) (say). As $t_n\rightarrow 0$ in the limit
$n\rightarrow\infty$, the solution gets contibutions from smaller and smaller
scales, inheriting the self- similarity of the
underlying equations (9). Consequently, the late time scaling  $x_f
\sim \eta^\nu$ could as well be derived starting from eq(9) for a
sufficiently large $n$, where $t_n=(-1)^n\ln\eta$. Note that there is
an ambiguity of sign depending upon $n$ being odd or even. However, this
only reveals the essential equivalence of (the scaling) behaviours in the
system's evolution both in the short, $\eta\rightarrow 0$ ($n$ odd), and long, 
$\eta\rightarrow \infty$ ($n$ even), time limit. We remark that, in case one
restricts all the variables $t_n$ in (0,1), as above, the
variable $t_0$ behaves only as a labeling parameter, the  flow of time is
solely indicated by $n$, the order of iteration. The iterated eq.(9) then
defines a  (discrete) map in the space of continuous functions.

\par Before interpreting the scaling, let us make a comparison of our
solution $x_i$ with $x_s$. Interestingly, eq(8) follows indeed from $x_s$
provided we inject time inversions via $t_n={1\over {1+t_{n+1}}}$,
$t_{n+1}\approx 0$ in the quadratic term in the exponential, collecting
together the dominant linear terms successively. As noted above, the
subdominant quadratic term finally drops out in the limit $n\rightarrow\infty$.
The time
inversions thus lead the system to evolve following the flow generated by
the intrinsic time sense. This proves our contention that the solution (8) is
an SL(2,R) extension of the standard solution. We note that {\em the solution
(8) gives essentially the late (short) time scaling of the standard
solution $x_s$ under inversions}. This means that the solution $x_s$, in a
physical application, represents the behaviour  of the system (given by
eq(1)) for moderately large $t$ ($\sim O(n)$, for moderate values of $n$).
However, for a sufficiently large $t$, the system's behaviour would slowly
deviate from  $x_s$, mimicking more and more the new solution $x_i$.

\par Finally, to interprete the scaling law, we note that the essential
linearity of $\ln x_i$  endows every self-similar replica of 
the evolving system, $\ln x_n$, the status of a `time keeper' (clock) which
can be used to record time $t_n$ after the $n$th iteration. The variable
$t_n$ can be defined as the $n$th generation Newtonian time for the
system (fluctuation) $x_n$. (Note that in every generation of linear evolution,
over the period (0,1), $t_n$ acts as an ordinary (nonrandom) variable).
However, as  discussed in Sec.3, $t_n$, having
constructed intrinsically from the $(n-1)$th generation interaction $v_{n-1}$,
corresponds to the intrinsic time relative to the $(n-1)$ generation
fluctuation $x_{n-1}$ (c.f., the relation of $t$ and $t_1$). The plethora
of time variables $t_n$, related to each other by inversions, as explained, 
are the linearised ramnants of the nonlinear, stochastic time, denoted $T(t)$, 
where the zeroth generation time $t$ is the ordinary Newtonian time. We note 
that once inversion is raised to a physically allowed mode of time flow, 
time as such becomes nonlinear. Our description above, however, identifies two 
simplified patterns of this nonlinear time (flow): i) the stochastic intrinsic 
flow $T$ following the ladders of the golden mean continued fraction ii) 
the extrinsic (linear) Newtonian flow $t$. However, the extrinsic time $t$ 
itself carries, thanks to (the possibility of) inversion(s),  
the seed of nonlinearity 
close to the instant $t=1$ (for instance). This fundamentally nonlinear 
behaviour of time could thus be represented succintlly as $T=t(T)$. Although, $T$ 
here may be considered to denote the `fully nonlinear' time, it is sufficient 
for our purpose to identify $T$ with the intrinsic time. Incidentally, we 
note that $T=t(T)\Rightarrow t=T(t)$, because of the interchangeability 
(relativity) of the extrinsic and intrinsic time at successive scales 
near $t=1$. Writing $\ln x_f \equiv\ln(x/\tilde x)=-T$, eq(4), rewritten
as 

$$
 {{dT(t)\over dt}=v(T)}\eqno(10)$$

 \noindent can now be considered as a renormalization 
group (RG) flow equation for the stochastic (nonlinear) time $T$. The
function $v(t)$ of eq(1), representing intrinsic (self-) interaction
of the system (equivalently, the nonlinear time) now assumes the
form of the RG $\beta$-function and  has the form $v(T)=\nu T=
\nu(1+\nu t(T))$ close to $t=(T=)1$, where $t(T)\approx 0$. 
(The factor $\nu$ could be considered 
as the measure of the strength of the time-time  self-interaction.
The free parameter $\lambda$ (c.f., eq(4)) then denotes
the strength of system's coupling with the nonlinear structure 
of time. Eq(10), being the defining equation of the nonlinear 
time, in particular, has $\lambda=1$). We note that nonlinear (stochastic)  
feature of time $T$ is revealed only close to $T=1$.
Assuming that time changes by pure translations only
even in the vicinity of $t=1$, we get the standard linear view of time. 
In this case a scale change is not allowed, so that $v(T)=1$.
Utilizing the SL(2,R) freedom near $t=1$ one gets a flow  of time which 
allows an evolving system to traverse finer scales, as explained in Secs.(2-3).
Clearly, eq(8) gives the unique nontrivial solution of the SL(2,R) RG
equation (10). The uniqueness, to within our definition of time inversion,
concerns only with the form of the solution. The randomness
in the scale-changing transition moments is likely to lead to very
different final states in long time scales for two systems with slightly
different initial conditions. In the next section, we present an ansatz 
for the nonlinear stochastic time $T$, and develope a method to compute the 
two-time correlation function $c(t)$ of a stochastic process obtained by replacing 
$t$ by $T$ in eq(1). Interestingly, our ansatz turns out to be an exact solution 
of eq(10) (c.f., Sec.7). As it turns out, the late (short) time power law of the 
solution $x_i$ correctly represents the same for the said correlation function. 
Consequently, the random transition moments $\tau_n$ (equivalently, the
periods $t_n$ of successive linear evolutions, randomness concentrating
near $t_n=0$ and $t_n=1$) could be considered as 
distributed with a probability distribution of a random walk process. The
correlation function $c(t)$ then gives the probability of not making a
transition at $t$, when it is known that no transition is made at $t=0$.
The nontrivial scaling now tells us that there is indeed a finite probability 
that the system does {\em make} a transition to a new scale in the long time limit, 
contradicting the linear time expectation that the probability should be 1. 
The asypmtotic power law of the correlation function
suggests that the late time evolution would resemble a Levy-like process [8]. 
This random walk scenario would be made more precise in Secs.5-7. 

\par We note also that the short time scaling $x_f \sim\eta^{\nu}$, 
on the other hand, could, 
be interpreted as the indicative of a fat fractal-like structure [7,10]
in the nonlinear time, with uncertainty exponent $\nu$, the golden mean.
Each point of the fattend time axis would thus be structured, equivalent 
to a Cantor set of dimension $\nu$. The non-zero uncertianty exponent now 
tells us that, contrary to the accepted notions, the precise determination
of an instant is almost impossible( even in the framework of the classical
mechanics). In fact, this could be inferred directly from eq(3), which reveals 
the fundamental role of golden mean in the definition of inversion. 
In the framework of nonlinear time, the problem of measuring a duration of unit 
length $t=1$ is equivalent to measuring a duration $\eta=\nu^{-1}, t=\nu\eta$, 
because $t$ is only a representative of the family of variables $t_n$.
However, the {\em irrationality} of the golden mean makes a precise measurement 
of the later impossible. Hence, one can at best conclude that $t\approx 1$, 
giving rise to {\em a fundamental limitation on measurability}. As an afterthought, 
this inference could even be  reached in the ordinary (linear time) framework. 
The problem of measuring $t=1$ can always be reduced to measuring a variable $\eta$ 
when $\eta=\nu^{-1}, t=\nu\eta$. The practical 
limitation of a precise determination of an irrational then translates to the above 
measurement limitation, giving rise to the width (uncertainty) necessary for 
inversion to materialize. This fact may then be considered as offering 
a (theoretical) justification to our assumption that time inversion is a physically 
viable mode of time flow. A relatively small error in measuring
the instant $t_1=1$ i.e., $t=\nu^{-1}$ (say), which is present unavoidably, now, 
 allows the system to explore
the inversion-induced intrinsic evolutionary path. In the process the initial error 
gets magnified, leading to the loss of final state predictability.  In this framework 
of fractal time, time inversion can thus be
interpreted as a consequence of the intrinsic uncertainty in ascertaining
if a moment near $t=1$ (say) is actually less or more than $t=1$.
We note finally that the model dependence in the late time scaling
exponent(=$\lambda\nu$) of the solution $x_i$ can be retrieved by reintroduing 
$\tilde t=\lambda t$ (cf. eq(4)).

\section{Stochastic versus quantum}

\par  Here, we present a framework to deal with an equation of the form eq(1) when 
time acquires a stochastic nature. The stochastic interpretation of time inversion 
allows one also to derive a Wheeler-Dewitt like equation for the fluctuating component 
$x_f$ of eq(1). In the light of the discussion of Sec.4, we envisage an ansatz for the 
stochastic (fractal) time, accommodating both an explicit inversion and the 
associated  randomness: $T(t)=(1+\mu \lambda(t)\tilde T(t))t$, where $0<\lambda(t)
<<1$ is a slowly varying function of the ordinary (Newtonian) time $t$, $\mu =
\pm 1$ is a (time independent) symmetric  random variable with $<\mu >=0,
<\mu ^2>=1$, $<,>$ being the statistical average, and $\tilde T(t)=T(t^{-1})$. 
A natural choice of $\lambda$ is, $\lambda(t) =\epsilon \nu ^nt$, $\nu$  being 
the golden mean number,
$n>0$ is a sufficiently large random integer, and $\epsilon = \epsilon (1/t)\sim 
O(1)$ could be an yet another slowly varying function which is constant over moderate 
scales, but may become `active' (i.e., large) near $t=0$. We disregard it 
in the following, but recall its presence in discussing short time scale structure of 
$T$ in Sec.7. We note that the arbitrariness 
in $n$ indicates the uncertainty in the actual moment injecting an inversion close 
to $t=1$ (say) and is used to model the uncertainty related to irrationality of $\nu$, 
whereas $\mu $ takes care of the definition of the inversion.
(Recall that inversion reflects an uncertainty in the neighbourhood 
 of $t=1$ (for instance). Thus, measuring an instant $t=1+ \tau, 
\tau>(\approx )0$ could very well end up with the result 
$t=1-\tau $ and vice versa.) For the sake of simlicity, we, however, choose $n$                                                                                          
to be a fixed interger (inversion is then assumed to materialize at a well-defined 
instant). We note, in particular that, $<T>=(1-\nu^{2n})^{-1}t$,(which encodes 
our discussion on the  limitation of measuability : although measurement 
of the instant $t=1$ (say) is exact in the context of the ordinary linear time, there is 
a uncertainty $O(\pm \nu^n)$ in the case of the fractal time $T$) and 
$T_1=\nu^nT=(1+\mu\nu^n t_1\tilde T_1)t_1, \tilde T_1=\nu^{-n}\tilde T\approx 1$, 
near $t=\nu^{-n}$. The implicit definition of $T$ is thus rescaling symmetric, 
exibiting its scale- free nature.
Further, $t\tilde T =T/t\sim O(1)$, by inversion symmetric 
nature of the ansatz (c.f., Sec.(7)). However, this behaviour might change 
in the limit  $t\rightarrow 0$ or $\infty$, because of large fluctuations 
generated due to  the activation of $\epsilon$ factors which are `mute' otherwise.  
Interestingly, the implicit definiton of $T$ turns out to satisfy the differential 
relation $dT={T\over t}dt$, which is valid for $t>0$. In the next section, we 
interprete this as a new class of stochastic nonlinear solution of an ordinary 
equation $t{dT\over dt}=T$ under inversion, indicating its relation with the 
measurement problem.

\par It now follows that the intrinsic scale dependent random flow of 
$\tilde T$ traverses the finer scale $O(\nu^n)$ ( c.f., Sec.3), as time $t$ flows to 
$O(\nu^{-n})$. In other words, 
contribution of the finer scale $\tilde T_1$ in a evolving system becomes significant 
$\sim O(1)$ at time $t\sim O(\nu^{-n})$. This random scale free flow of fractal time 
$T$ is responsible for approximate self-similarity of the system's 
evolution. To justify this, we note that a time dependent equation of the form eq(1) 
can be written generically as the stochastic equation 

$$
 {{dx\over dt}=(1+\mu\nu^n t\tilde T)h(T)\,x}\eqno(11)$$

\noindent where $dT={T\over t}\,dt$. Note that, close to an instant $t$, ${T\over t}h(T)
\approx(1+\mu\nu^nt \tilde T)[h(t)+\mu \nu^n T {dh\over dt}]= h(t)+[\mu\nu^nt\tilde Th(t) 
+\mu\nu^n (t\tilde T)T{dh\over dt}]$, so that 
writing $x=e^{\int h(t)dt}\bar x_f, \,(x(0)=1)$, eq(11) reduces to 

$$
 {t{d\bar x_f\over dt}=[\nu^{2n}Th_1(T_1)+\zeta (t)]\,
\bar x_f}\eqno(12)$$

\noindent where $h_1(T_1)=h_1(t_1)+\mu\nu^n T_1 {dh\over dt_1}$, close to 
the instant  $t_1(=\nu^n \ln t)=0$, and $\zeta(t)=\nu^{2n}T\{(\mu\nu^{-n} h(t)\\ 
-h_1(t_1))+\mu(\nu^{-n}T{dh\over dt}-\nu^n T_1{dh_1\over dt_1})\}$, acts as a 
`noise' at the scale changing transition point. The source of this noise is the 
mismatch of the boundary values of the function $h(t)$, because of the $\mu$ factors, 
although  $h(t)=h_1(t_1),\,{dh\over dt}=\nu^n{dh_1\over dt_1}$, for $t=1+\bar t$ (say).
It follows from the relations $<\mu T>=\nu^{2n}<T>,\,<\mu T^2>=2\nu^{2n}<T^2>\approx 
2\nu^{2n} <T>^2$, which can be proved using the representation $T=(\Sigma \mu^r\lambda^r)t$ 
(obtained by repeated applications of the ansatz over itself), that $<\zeta >=0$, upto 
order $O(\nu^{2n})$. We remark, in passing, that, in this fromalism, 
the expectation of the integrated residual noise is expected to remain finite in longer 
time scales. Further elaboration of this point would be considered separately.

\par It now follows that eq(12) is  self-similar to eq(11), but for the zero mean 
noise term. (Eq(11) is, in fact, form invarinat under rescalings when 
the nonrandom mean evolution of the system is removed.) However, removing the noise term 
by writing $\ln \bar x_f=\int \zeta (t)t^{-1}dt+\ln x_f$, we finally get the 
correct self-similar equation for the fluctuation $x_f$:

$$
 {-{dx_f\over dt_1}=(1+\mu\nu^n {T_1\over t_1})h_1(\tilde T_1)\,x_f}\eqno(13)$$

\noindent In the derivation of this equation, via eq(12), we make use of  $t_1\tilde T_1
=t\tilde T =T/t, \, T_1=\nu^n T$ whenever necessary. The (-) sign in the left hand side, 
and other symmetric changes in the right hand side are consequences of inversion near 
$t=1$. We note that the logarithmic derivative in eq(12) ensures the initial 
condition $x_f(0)=1$ at $t_1=0$ for eq(13).  To explain eq(13), we note, following Sec.3, 
that the system $x$ is evolved till $t\approx 1$, when the mean (nonrandom 
part of the) evolution is removed. An inversion near $t=1$ then induces a scale changing 
process infusing the fluctuating system with a noise, generated due to `boundary effects'.
Once the noise is  filtered out, the fluctuating component $x_f$ is found to evolve 
following a reduced equation which is self-similar to the original equation till $t_1\sim 1$ 
(our approximations break down beyond $t_1=1$). Near the epoch $t_1=1$, the system again 
gets ready to accommodate another scale changing process $t_1\rightarrow t_2$, and so on for 
the subsequent evolution. This completes our derivation of self-similarity of eq(1) in 
the framework of the stochastic time.
\par  We note that removal of the total mean evolution by the ansatz $x=e^{\int <Th(T)>dt}
\bar x_f$, instead, would have resulted the equation

$$
 {t{d\bar x_f\over dt}=(Th(T)-<Th(T)>)\bar x_f}\eqno(14)$$

\noindent so that $<{d\bar x_f\over dt}>=0$. This is analogous to the Wheeler-
Dewitt-like  equation in quantal evolution, indicating the reparamtrization 
invarinace of geometric phase [7]. In the present classical context, this could be 
interpreted as the reparametrization invarince of the fluctuating evolution, 
due to time inversion. However, $<Th(T)>=h(t)+O(\nu^{2n})$, so that the above 
equality is only approximate, in all practical purposes; whence the self- similar
 eq(13) of the fluctuation could be  retrieved (c.f., Sec.4, also see Sec.7).  The 
reparametrization invarince of eq(14) thus offers a justification in 
calling the variable $t_1$ an intrinsic time. A deeper analysis of this 
fractal time approach in (quantum) general relativity needs separate investigations.

\section{$1/f$ Spectrum and Applications}

\par  We now show that the two-point correlation function of 
the fluctuating system  has a late time power law form, resembling a Levy 
like process [8]. Let $c(t)=<\bar x(t)\bar x (0)>, \bar x=e^{-\int h(t)\,dt}x$ denote the 
correlation function of the evolving system $x $ in eq(11). Here, $<,>$ 
denotes the statistical average 
over an ensemble of fluctuating solutions (for all possible realizations of the 
scale changing transition moments, in the terminology of Sec.4) with a common 
initial condition. Because of self-similarity of eqs.(11) and (13), the fluctuation
$x_f$ and the original system $x$ have identical structures. The late time asmyptotic 
form of the correlation function $c(t)=<\bar x(t)>$ (for the initial condition $\bar x(0)=
x(0)=1$) can be easily obtained from eq(11)
by first estimating it near $t=1$ and then taking the limit $\epsilon \rightarrow 0$
of a slowness parameter $\epsilon $ (a suitable power of $\nu$, and not to be confused 
with one in the previous section), which occurs naturally  
as a rescaling parameter of the time variable:$t\rightarrow \epsilon t$ (c.f., Sec.4), 
in the present scenario.  Clearly, in the limit $t=1$, eq(11) (via eq(12), neglecting 
higher order terms) reduces to 
$$
 {t{d\bar x\over dt}=\mu\nu^n t\tilde T h_1\,\bar x}\eqno(15)$$

\noindent so that $d<\bar x>=\nu^{2n}h_1 <\bar x> t^{-1}dt,\,h(1)=h_1$, when we use 
$<\mu\tilde T(t)>=\nu^n<\tilde T>=\nu^n t^{-1}$, upto $O(\nu^n)$. One thus gets an 
ordinary differential equation, analogous to eq(4), for the correlation function 
$c(t)$. Following Sec.3, this could be solved near $t=1$, accommodating a time 
inversion: $t^{-1}dt\rightarrow -t^{-1}\,dt$. We thus get $c(t)\sim t^{-\gamma },\, 
\gamma =\nu^{2n}\,h_1$, near 
$t=1$. The late time asymptotic form $c(t)$ is now obtained by taking $\epsilon 
\rightarrow 0$ and renormalizing it by a vanishingly small factor: $C(t)=
O(\epsilon ^\gamma )c(t)\sim  t^{-\gamma }$. Necessity of renormalizing $c(t)$ by 
a small factor does not
jeopardy its physical relevance. In fact, the relevant equation of $c(t)$, (after an 
inversion) near $t=1$, $dc=-\gamma  t^{-1} c\, dt$, being rescaling symmetric, is valid 
even for large $t$. However, the possible presence of a small factor might be responsible 
for not having any inkling so far about the relevance and implications of time inversion 
in the context of eq(1). Interestingly, we note here that the asymptotic form of $c(t)$ 
mimics the late time 
behaviour of the new solution $x_i$ in eq(8). (The power law form $\sim \eta^{-\nu}$, 
in the paragraph below eq(8), is written in the log-scale of $t_1$, instead of $t$, 
as in here. Expressed in the log-scale of $t$ the exponent is $\nu^2$. The role of $n$ 
in the exponent of $c(t)$ is explained in the next section.) Consequently, 
the solution $x_i$ could be  identified as the correlation 
function for the stochastic process, eq(11). Finally, the power spectrum 
$S(f)$ of the inverse power law correlation function $c(t)$, $S(f)=2\int_0^\infty c(t)
\cos(2\pi ft)\, dt$, is known to diverge [3,9] with a power law tail 
$\sim 1/f^{1-\gamma }$, in low frequency limit. 

\par In the same spirit, we can also make an estimate of the short time scaling of 
the fractal time $T$. Recalling $\epsilon =1+{1\over 2}\epsilon_0 t^{-1}$, which appears 
with $\nu^n$ as a constant ($\approx 1$) factor for $\epsilon_0<<1$ (a factor of 1/2 
is included for convenience), and can now play 
a role to generate small scale time variations, $<T>$ can be expressed as 
$<T^\prime>=[(1-\nu^{2n})-\nu^{2n}\bar t]^{-1}$, where $\bar t=\epsilon_0 t^{-1}<<1$, 
and $T^\prime=t^{-1}T$,
so that neglecting  $\nu^{2n}$ compared to 1, we get $<T^\prime>=(1-\nu^{2n}\bar t)^{-1} 
=\tilde t^{\nu^{2n}}, \tilde t=1+\bar t$. Let us note further that, as remarked in Sec.5, 
$<T^\prime>$ diverges at $t=0$ because of $\epsilon $. However, away from $t=0$, there 
always exits an interval of small $t:\, \epsilon_0<<t<<1$, when the above estimate
of $<T^\prime>$  is well defined. Using $\epsilon_0$ as a cut off, one can treat 
$T^\prime(0)(= T^\prime(\epsilon_0))=1+O(1)\mu\nu^n$, as a finite time independent random 
variable.  The relevant correlation function  $ c^\prime(t)=<T^\prime(t)T^\prime(0)>
\sim <T^\prime(t)>$, thus is given by $c^\prime(t)\sim \tilde t^{\nu^{2n}}, \tilde t
\approx 1$. Following above arguments, we conclude that 
$ c^\prime(t)\sim  t^{\nu^{2n}}$ for small $t$. Again this power law mimics the power law 
obtained in the last paragraph of Sec.4. We remark that the equivalence of the scalings of 
$c(t)$ and $c^\prime(t)$ is indicative of the fact that the `a priori' nature of Newtonian 
time gets blurred in longer time scales; system's evolution is more accurately 
described by a class of intrinsic time variables, generated by the scale- free 
fluctuations, as inherited by the system from the fractal nature of time.

\par From the above discussions, it appears very natural to believe that observations 
of $1/f$-like spectra in many natural phenomena must arise, at least partially, from 
the {generic} principle of time inversion induced stochastic fluctuations. We note 
that $1/f$ spectra in time series records, for instance, of quasar light intensity 
fluctuations, river (ocean) water level fluctuations [3], temperature (voltage) 
fluctuations under a steady current [9], etc could in fact be understood  as  
generic consequences of time inversion. One can always concieve $x$ in eq(1), as 
representing the relevant flucutating variable, the rate of change of whose 
variation is supposed to be proportional to the system variable itself, where 
the `proportionality constant' is a slowly varying function $h(t)$ of time. In general, 
$h(t)=h(t,x)$ may be a nonlinear function of the system variable $x$ and other external 
influences. Here, we disregard explicit modelling of external influences, and assume 
that the time variation of $h(t)$ arises purely from the nonlinear influences of the 
fractal time $T$. The relevant equation, eq(1), then assumes the form of the stochastic 
equation (11), giving rise to the generic divergence in the corresponding spectrum. 
Numerical fits to time series data should be useful in  estimating the model dependent 
index $n$ of the spectrum.  

\section{Fractal time and Measurement limitation}

\par Here, we present some salient features of our ansatz for the fractal 
time in Sec.5. In fact, we show that the ansatz constitutes a new class of stochastic 
solutions  to the simplest linear differential equation ${dx\over dt}=x$ (c.f., eq(10)). 
Let us rewrite the ansatz in the form $T(t)=(1+ \lambda\, t\,\tilde T(t))\,t$, where 
$\tilde T(t)=T(1/t)$ and $\lambda $ may be an almost constant  slowly varying 
function of $t$. To begin with  we disregard any explicit randomness in $\lambda$.
By symmetry, both $T/t$ and $t\tilde T$ satisfy coupled equations
of the form $x=1+\lambda y$ and $y=1+\lambda x$, hence $T(t)/t=t\tilde T$ for all $t$ 
(when time variation of $\lambda $ is disregarded). 
We assume that $T(t)$ is continuously differentiable except possibly at $t=0$ and $\infty$. 
Noting that ${d\tilde T\over dt}=-t^{-2}{d\tilde T\over dt^{-1}}$, we get 
${dT\over dt}= (1+\lambda t\tilde T)+t\lambda \tilde T -\lambda {d\tilde T\over dt^{-1}}$, 
so that ,$({dT\over dt}-{T\over t})+ \lambda ({d\tilde T\over dt^{-1}}-t\tilde T)=0$. It 
thus follows, $\lambda $ being an arbitrary scale factor,  that 
$$
{t{dT\over dt}=T}\eqno(16)$$ 

\noindent as asserted. The nontrivial role of inversion in obtaining the result needs 
to be emphasised. The above equation is not satisfied for an ansatz with $T$, 
replacing $T$, in the braket. It also follows that the relevant class of functions 
$T(t)$ must satisfy the condition ${dT\over dt}={d\tilde T\over dt^{-1}}$.
We note that the equation is exact for a constant 
$\lambda$. However, the algebraic constraint $x=1+\lambda \,x$ implies 
$x=(1-\lambda )^{-1}$, which means $T=(1-\lambda )^{-1}t$. Consequently, for rational 
values of $\lambda $, when an exact evaluation of the scale factor is permissible, our
ansatz, being the trivial (standard) solution of the above equation, fails to yield 
any new solution. However, for an irrational value of $\lambda $, the standard linear 
solution is valid only approximately $T\approx (1-\lambda )^{-1}t$, in any physical 
application, which is a direct consequence of the measurement uncertainty discussed 
in Sec.4. As stated already, we (partially ) model this uncertainty by introducing a 
random parameter, 
$\lambda \rightarrow \mu \lambda $. In this case, the algebraic constraint assumes 
the status of a stochastic equation  $x=1+\mu\,\lambda \,x$, so that the standard 
solution is obtained only in the mean: $<x>=1+\lambda ^2<x>$, $<T>=(1-\lambda ^2)^{-1}t$. 
The nonlinear stochastic function $T(t)=(1+ \mu\,\lambda\, t\,\tilde T(t))\,t$, then 
represents a nontrivial solution of eq(16). Finally, to fix the 
scaling parameter $\lambda $, we concieve the ideal situation of perfect time 
measurement $<T>=1$ at $t=\nu$  leading to the value $\lambda =\nu$.  
In the present framework, this ideal time measurement is obviously precluded in natural 
phenomena, making a way to fluctuations over many time scales and complex structures.	 
We remark that the choice of $\nu$ here is motivated by the SL(2,R) representation, eq(3), 
and the fact that the convergence rate of the golden mean approximants is the slowest 
possible, leading to the slowest ever rate of the intrinsic flow (c.f., Sec.3). 
For a rational value of $\lambda$, however, the scope of a nontrivial inversion 
is eliminated. The inversion in the implicit definition of $T$ then corresponds to 
the ordinary time reversal symmetry only (c.f., last paragraph of Sec.2).
Mathematically, $\lambda $ can of course be any irrational number in $0<\lambda<1$.

\par To explore the nontrivial role of inversion (as defined in Sec.2) in the 
above discussion further, let us recall that the inversely related moments 
$t_{-}$ and $t_{+}$, with intrinsic uncertainties, are defined by $t_{-}t_{+} 
=1+\delta $, where, $\delta $ is an $O(\bar t^2)$ random variable. In the present 
fractal time framework, one can model this random, nonlinear behaviour by 
the definition $t_{+}=T(t_{-}^{-1})=(1+ \lambda\, t^{-1}_{-} T(t_{-}))\,t^{-1}_{-}$, 
so that $t_{-}t_{+}= [1+\lambda t^{-1}_{-} T(t_{-})]$. One can now verify 
easily that $T=t_{-}(1+\lambda t_{-}t_{+}) $ represents a nontrivial solution 
of eq(16). Clearly, the small (nonrandom part of ) parameter $\lambda$ avoids 
any clash with standard observations at moderate scales. In fact, $T=t$, till 
$t\sim O(\lambda^{-{1\over 2}})$, influences of random multiple scales would be 
felt only in the longer time scales. The explicit form of a generic nontrivial 
solution indicating all the scales (analogous to the Weierstrass function) is 
still missing.

\par To understand the role of the index $n$ in the previous two sections, we now 
show that the fractal time $T(t)$ in fact represents a more general mulitiplicative 
process. Denoting $T_n=(1+\mu\nu^n t_n\tilde T_n)t_n$, let us define the general 
fractal time T by  the multiplicative process, $T/t=({\chi \over t})(t\tilde\chi)
=({\chi \over t})^2$ (by inversion symmetry), where $\ln \chi (t)=\Sigma\,\nu^{n+1}
\ln T_n(t_n)$. It is now easy to check that $T$ satisfies the equation $t\,dT=T\, dt$, 
when each of $T_n$ solves $t_n\,dT_n=T_n\, dt_n$. To interprete $T$, we may imagine 
that a grand complex process represented by $T$ is materialied when the sequence of 
sub-processes $T_n$ is successfully materialized with respective probability 
of success $\nu^{n+1}$. The process is materialized (survived) at longer time scales 
$t\sim O(\nu^{-n})$ only with lesser probability. The short time correlation function 
$C(t)$ of the grand $T$ is obtained from  the defining relation by taking expectation 
values: $\ln <T(t)t^{-1}>=2\Sigma\,\nu^{n+1}\ln <T_nt_n^{-1}>$, so that $C(t)\sim 
t^{\nu^2}$. The correlation function $c(t)$ obtained in Sec.6 thus corresponds to the 
$n$th level fractal time $T_n$. The exponent of $C(t)$ correctly recovers the same 
obtained in Sec.4. Our method of solving eq(1) in Sec.3, giving rise to the solution 
$x_i$ in eq(8) thus directly leads to the correlation function $C(t)$ of the grand 
fractal time $T$.  We note that any natural process, being manifested only over a finite 
period of time, would fail to display the grand fractal time exponent $\nu^2$, and  would 
at most equal to a finite sum of powers of $\nu^2$, that too for large $n$.

\section{Conclusion}
Our study on the structure of time reveals a number of surprises. Time may 
indeed flow by inversions, leading to multiscale stochastic 
evolution for a linear system. Inversions ascribe a stochastic fractal like 
structure to time itself. We discuss two methods in uncovering the late time 
stochastic feature of a linear system. Repeated applications of inversion 
close to well defined scale dependent moments of the form $t_n=1$, give rise 
to a new solution, self-similar over multiple scales, which can be interpreted 
as the two time correlation function of the corresponding stochastic equation, 
when the ordinary time variable $t$ is replaced by a fractal time $T(t)$. We present 
an ansatz for the fractal time which turns out to represent a new class of fractal 
solutions to the simplest linear differential equation. As a consequence, an 
ordinary linear system attains the status of a stochastic process. Our 
ansatz for fractal time provides a framework to compute the correlation 
function and the corresponding power spectrum of the process. We show that 
such a process would generically enjoy multiscale self-similarity leading 
to $1/f$ spectrum. We discuss intricate relations of the definition of 
time inversion, fractal time, irrationality of the golden mean and the 
fundamental uncertainty in measurability of a duration. Finally, we 
have discussed the relevance of intrinsic time in the present formalism  drawing 
interesting analogies with the Wheeler-Dewitt equation. Our analysis show that 
the distinction between time and the system variable tends to get obliterated in 
longer time scales. We close with the remark 
that applications of the present fractal time approach is likely to yield new 
understanding mostly in the short and long time scales of dynamical systems.
The scope of its applicability in quantum field theories also need not be 
overemphasised.

\section*{Acknowledgement}
It is a pleasure to thank Inter University Centre for Astronomy and 
Astrophysics , Pune, India for hospitality under its Associateship 
programme where this work was initiated.

\end{document}